\documentclass[twocolumn,showpacs,preprintnumbers,amsmath,amssymb]{revtex4}

\usepackage{graphicx}
\usepackage{dcolumn}
\usepackage{bm}

\begin{document}

\preprint{}

\title{Photonic spin Hall effect in metasurfaces with rotational symmetry breaking}
\author{Yachao Liu$^{1}$}
\author{Xiaohui Ling$^{2}$}
\author{Xunong Yi$^{2}$}
\author{Xinxing Zhou$^{1}$}
\author{Shizhen Chen$^{1}$}
\author{Yougang Ke$^{1}$}
\author{Hailu Luo$^{1}$}\email{hailuluo@hnu.edu.cn}
\author{Shuangchun Wen$^{1}$}\email{scwen@hnu.edu.cn}
\affiliation{$^1$ Laboratory for Spin Photonics, College of Physics
and Microelectronic Science, Hunan University, Changsha 410082,
China\\$^2$SZU-NUS Collaborative Innovation Center for
Optoelectronic Science Technology, and Key Laboratory of
Optoelectronic Devices and Systems of Ministry of Education and
Guangdong Province, Shenzhen University, Shenzhen 518060, China}
\date{\today}

\begin{abstract}
Observation of photonic spin Hall effect (SHE) in dielectric-based
metasurfaces with rotational symmetry breaking is presented. We find
that the spin-dependent splitting is a unique angular splitting in
the real position space, and is attributed to the space-variant
Pancharatnam-Berry phase (PB). Breaking the rotational symmetry of
the PB phase by misalignment of the central axes of the incident
beam and the metasurface, the spin-dependent shift is observable. We
show that the spin-dependent shift can be enhanced by increasing the
rotation rate of the metasurface, so the metasurface provides a
great flexibility in the manipulation of photonic SHE.
\end{abstract}

\pacs{42.25.-p, 42.79.-e, 41.20.Jb}
\keywords{photonic spin Hall effect, Pancharatnam-Berry phase,
metasurface}

\maketitle

\section{Introduction}\label{SecI}
In quantum mechanics, spin is an intrinsic angular momentum, which
is the inherent nature of elementary particles. Electronic spin Hall
effect (SHE) is a transport phenomenon in which an electric field
applied to spin particles results in a spin-dependent shift
perpendicular to the electric field
direction~\cite{Murakami2003,Sinova2004,Wunderlich2005}. Photons can
be assigned with two opposite spin states with the spin axes
parallel and anti-parallel to the wave vector, which correspond to
left- and right-circular polarizations, respectively. Photonic SHE,
which is generally believed to be a result of topological spin-orbit
interaction~\cite{Hosten2008,Bliokh2008I}, is just the photonic
counterpart of the SHE in electronic
system~\cite{Onoda2004,Bliokh2004,Bliokh2006,Duval2006}. The
spin-orbit interaction describes the coupling between the spin
(circular polarizations) and orbital degrees of freedom of photons,
which is the signature of two types of geometric phases: the
Rytov-Vladimirskii-Berry (RVB) phase and the Pancharatnam-Berry (PB)
phase. The former is related to the evolution of the propagation
direction of light~\cite{Onoda2004,Bliokh2006,Hosten2008} and the
latter to the manipulation with the polarization state of
light~\cite{Bliokh2008I,Bliokh2008II}.

When a linearly polarized paraxial light beam impinges obliquely
upon an interface between two different media, the spin-dependent
splitting in real position space generates, which is associated with
the RVB phase. This photonic SHE draws considerable attention in
recent years due to its potential applications in precision
metrology~\cite{Zhou2012I,Zhou2012II} and spin-based
photonics~\cite{Hosten2008,Shitrit2013}. However, the shift induced
by the photonic SHE is very tiny, usually with the order of a
fraction of the wavelength. Weak measurement is a sensitive method
to detect the photonic
SHE~\cite{Hosten2008,Qin2009,Luo2011,Gorodetski2012}, which enlarges
the original tiny displacement with preselection and postselection
technology. Recently, Yin \textit{et. al.} demonstrated
experimentally a spin-dependent splitting in a thin metasurface with
designed in-plane phase discontinuity at the wavelength scale, which
is related to the RVB phase~\cite{Yin2013}. Within their scheme,
opposite spin states accumulated at the opposite edge of the beam
and constructed a spin-dependent splitting perpendicular to the
designed phase gradient.

While for an inhomogeneous anisotropic medium (e.g., liquid crystal
q plate), the light beam can acquire the PB phase. The inhomogeneous
anisotropic medium is usually be presented with rotational symmetry,
which is expected to produce vortex beams or vector beams with great
potential in microparticle manipulation and quantum information
~\cite{Bomzon2001,Marrucci2006,Zhao2013}. With rotational symmetry
of the medium, the produced PB phase would also be locally varying
and forms a geometric phase gradient in the azimuthal direction.
Actually, the PB phase is spin-dependent, that is, the spin of the
incident photons determines the sign of the phase.  However, if the
incident light involves both spin components, these components would
always superpose at the corresponding position due to the rotational
symmetry, although they should have acquired just opposite PB phase.
Breaking the rotational symmetry, it is likely to observe a
spin-dependent splitting of light, i.e., photonic SHE, due to the
geometric phase gradient. It has been reported in a one-dimensional
plasmonic chain, which can be viewed as breaking a rotationally
symmetric structure from polar coordinate to the Cartesian
coordinate~\cite{Shitrit2011}. More recently, spin-dependent
splitting in metasurfaces has been observed when the spatial
inversion symmetry is violated~\cite{Shitrit2013}. The resulted
photonic SHE is large enough for direct measurements, in contrast
with the indirect detecting technology using the weak measurement.

In this work, we report the observation of photonic SHE in
dielectric-based metasurfaces with rotational symmetry in which we
break the rotational symmetry of PB phase by means of misaligning
the central axes between the light beam and the metasurfaces. In
recent years, metasurfaces is gaining a reputation for introducing a
designed geometric phase~\cite{Shitrit2013,Yu2011,Kildishev2013}.
The metasurfaces applied in this work are fabricated by etching
locally varying grooves in a silica glass. As the optical dimension
of the grooves is much less than the working wavelength, it creates
a metasurface with inhomogeneous anisotropy due to the form
birefringence~\cite{Beresna2011}. It produces a spin-dependent
angular splitting of light in the real position space as compared
with the parallel splitting related to the RVB phase. Suitably
engineering the metasurface geometry, we can achieve any desirable
PB phase, and thereby the spin-dependent splitting of light.

\section{Theoretical model}\label{SecII}

\begin{figure}
\includegraphics[width=8cm]{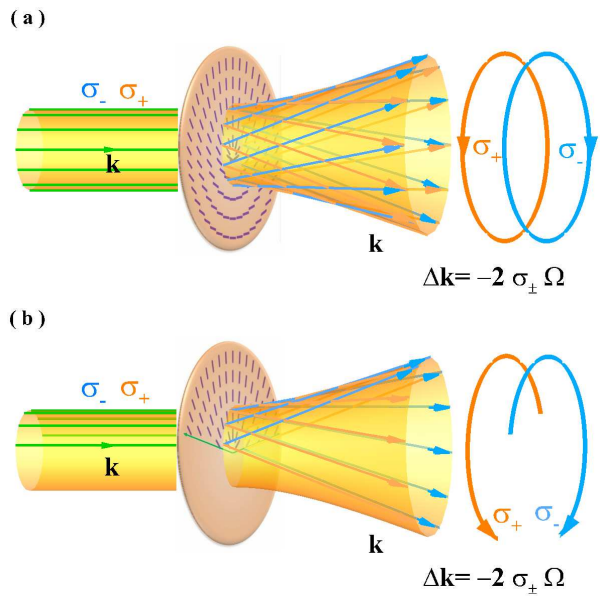}
\caption{\label{Fig1} (Color online) The contrast between the
metasurfaces with rotational symmetry (a) and with rotational
symmetry breaking (b). The rotational symmetry can be broken by
designing the metasurface structure, which just maintain part of the
inner structure in comparing with the metasurface with rotational
symmetry. The notations $\sigma_{+}$ and $\sigma_{-}$ represent
left- and right-spin states respectively, and $\Omega$ is rotation
rate of metasurface structure.}
\end{figure}

In order to get insight into the physical mechanism of the
metasurface, Jones calculus is performed here to analyze the
polarization and phase transformation. The Jones matrix of the
metasurface with inhomogeneous local optical axes and constant
retardation $\pi$ can be written as ~\cite{Liu2014}
\begin{equation}
\textbf{J}=\left(\begin{array}{ccc}
\cos{2\phi} & \sin{2\phi} \\
\sin{2\phi} & -\cos{2\phi}
\end{array} \right)\label{Matrix},
\end{equation}
where $\phi$ is the local optical axis direction. When a
right-circular ($|R\rangle$) or left-circular ($|L\rangle$)
polarization beam normally impinges into the metasurface, the output
state can be calculated as
\begin{equation}
|E_{out}\rangle=\textbf{J}|R\rangle=\exp(-i2\phi)|L\rangle\label{2},
\end{equation}
\begin{equation}
|E_{out}\rangle=\textbf{J}|L\rangle=\exp(i2\phi)|R\rangle\label{3},
\end{equation}
where $|R\rangle=(1,-i)^{T}/\sqrt{2}$ and
$|L\rangle=(1,i)^{T}/\sqrt{2}$ represent the right- and
left-circular polarization, respectively. It is clear that an
additional phase factor $2\phi$ is induced in this process and its
sign is spin-dependent. Deriving from the conversion of polarization
states, obviously, this additional phase is the PB phase in nature
and depends only on the orientation of local optical axes. As the
local optical axes varying with location, the passing beam will
obtain a position-dependent PB phase.

When the metasurface be engineered with rotational symmetry
[Fig.~\ref{Fig1}(a)], the induced PB phase would form a spiral beam
wavefront because of the linear rotation of the local optical axes
in the azimuthal direction. In this way, both right- and
left-circular polarizations (spin states) can acquire a helical
phase. For different spin states, the helical phases are opposite,
which would bent the wave vectors along the spin-dependent phase
gradient as shown in Fig.~\ref{Fig1}. However, the wave vectors of
two spin states still overlap with each other in the real position
space, so that there is no spin-dependent splitting in this
situation [Fig.~\ref{Fig1}(a)]. Hence, we manage to keep the
tendency of deflection but separate the opposite spin states. It is
realized by scrubbing part of the inner structure of metasurface as
shown in Fig.~\ref{Fig1}(b). Removing part of the inner structure
will break the rotational symmetry of metasurface so as to make the
separation of spin states observable in the real position space.

Note that the geometric phase gradient in the azimuthal direction is
proportional to the rotation rate of local optical axes, which is
deduced from Eqs.~(\ref{2}) and ~(\ref{3}). When we refer to a small
part of the metasurface as shown in Fig.~\ref{Fig1}(b), the phase
gradient can be approximately regarded as in the horizontal
direction. Thus, the constructed geometric phase gradient can be
related to the rotation rate of the metasurface in the form:
\begin{equation}
\nabla_{x}\Phi=d\Phi/dx=\frac{\sigma_{\pm}2d\phi}{dx}=2\sigma_{\pm}\Omega\label{PBPG}.
\end{equation}
where $\sigma_{\pm}$ is the incident spin state, $\Omega$ denotes
the rotation of local optical axes in a unit length $d\phi/dx$,
$\Phi$ is the induced PB phase.

As this derivation, when we refer to the normal incidence, the
emerged beam should be diffracted due to the PB phase gradient along
the metasurface. The transmission angle is proportional to the phase
gradient in the small-angle approximation:
\begin{equation}
\theta_{t}=\frac{\lambda_{0}}{2\pi}\nabla_{x}\Phi\label{SDT},
\end{equation}
where $\lambda_{0}$ is the vacuum wavelength.

\begin{figure}
\includegraphics[width=8cm]{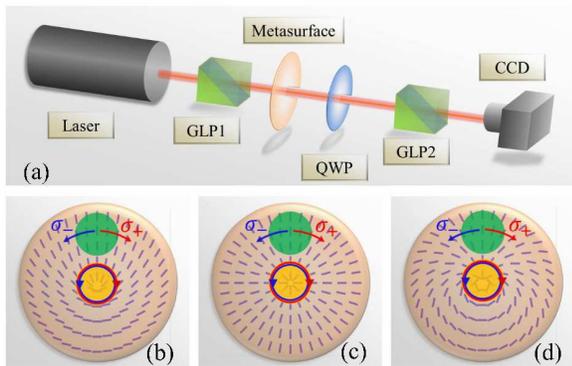}
\caption{\label{Fig2} (Color online)(a) The experiment setup to
observe the photonic SHE in metasurface with rotational symmetry and
with rotational symmetry breaking. The light source is a 21 mW
linearly polarized He-Ne laser (632.8nm, Thorlabs HNL210L-EC). GLP,
Glan laser polarizer; HWP, half waveplate; QWP, quarter waveplate;
CCD, charge-coupled device (Coherent LaserCam HR). [(b), (c), and
(d)] are metasurfaces with three different rotation rates employed
in our experiment. Misalignment of the central axes between the
incident beam and the metasurface would separate the overlapped spin
states of light by removing the rotational symmetry of PB phase.
Yellow circles mark the incident position for beam to keep the
rotational symmetry of metasurface and green circles are the
situations with rotational symmetry breaking. Red and blue arrows
(notations $\sigma_{+}$ and $\sigma_{-}$) are the wave vector
components of opposite spin states in the plane of metasurface.}
\end{figure}

According to the Eqs.~(\ref{PBPG}) and ~(\ref{SDT}), we find that
the transmission angle $\theta_{t}$ should be opposite in sign for
different incident spin states. This spin-dependent angle $\theta_t$
and the spin displacement $d$ can be represented by
\begin{equation}
\theta_t=\frac{2\sigma_{\pm}\Omega}{k_{0}},~~~d=\frac{2\sigma_{\pm}\Omega}{k_{0}}z,\label{SDS}
\end{equation}
where $k_0=2\pi/\lambda_0$. These results show an angular splitting
in the real position space. The intensity splitting can be induced
after the light propagating to the far field.

\begin{figure}
\includegraphics[width=8cm]{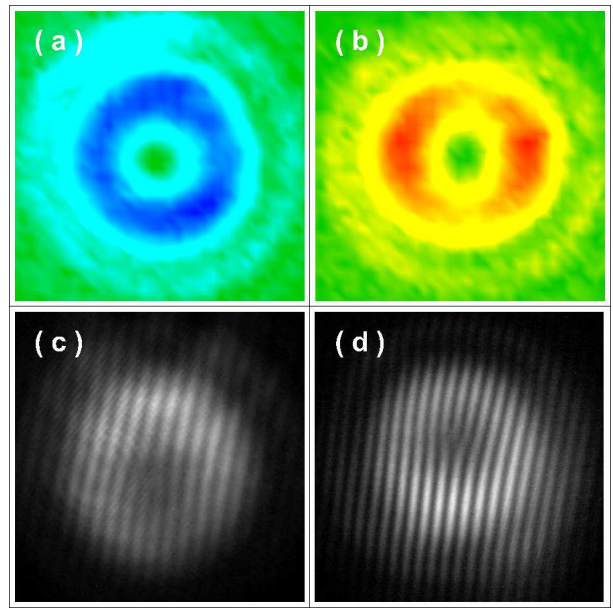}
\caption{\label{Fig3} (Color online) (a) and (b) The intensity
pattern of emerging beams when a circularly polarized beam is
impinging on the metasurface with rotational symmetry. Red and blue
represent the right- and left-circular polarizations, respectively.
(c) and (d) Intensity pattern obtained from the interference between
a reference beam and the optical vortices obtained in our
experiment. The dislocation of the fringe pattern indicates the
presence of a vertex phase, whose sign of topological charge depends
on its incident spin. }
\end{figure}

It is clear that the photonic SHE usually refers to a transverse
spin-dependent splitting, when a spatially confined light is
reflected or refracted at an interface. This transverse splitting is
generally independent of the propagation distance in real position
space. However, the splitting in our scheme is totally different
with these ordinary situations. By introducing a spin-dependent PB
phase gradient in the metasurface, a spin-dependent angular
splitting is demonstrated. The splitting angle is proportional to
the phase gradient along the surface, and the spin displacement will
increase upon propagation. For an appropriate designed rotation
rate, the spin-dependent splitting of intensity would be enhanced
for a direct observation.

\section{Experimental observation}\label{SecIII}
We implement an experiment to demonstrate the angular splitting and
the photonic SHE [Fig.~\ref{Fig2}(a)]. A He-Ne laser is applied as
the light source. Beam passing through the Glan polarizer impinges
into the metasurface. A quarter waveplate and a second polarizer
cooperate with the charge couple device (CCD) to record the spin
distribution of the output beams. There are three different
metasurfaces applied in our experiment, which are schematically
demonstrated in Figs.~\ref{Fig2}(b)-\ref{Fig2}(d), respectively. The
three metasurfaces applied in our experiment are designed with
different rotation rates. By etching grooves with appropriate
directions and geometrical parameters, we can achieve effective
birefringence in the metasurface (Altechna). The nano-grooves in the
metasurface with uniform phase retardation $\pi$ can reverse the
circular polarizations and introduce the desired PB phase.

\begin{figure}
\includegraphics[width=8cm]{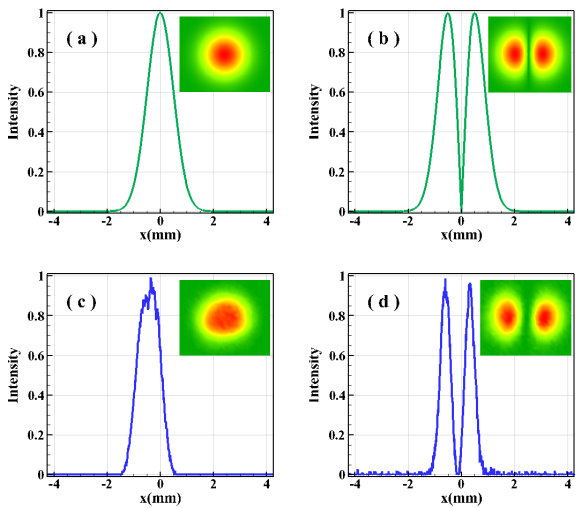}
\caption{\label{Fig4} (Color online) Intensity of the
cross-polarized component after the beam propagating through the
metasurface. The input states are chosen as horizontal polarization
(left column) and the output states is vertical polarization (right
column). [(a),(b)] Theoretical results; [(c),(d)] Experimental
results. Insets present a full view of the intensity distributions.}
\end{figure}

When a linearly polarized beam incident upon the metasurface along
the central axis, where the incident position is marked by the
yellow circles in Figs.~\ref{Fig2}(b)-\ref{Fig2}(d), the metasurface
applies rotationally symmetric response to the beam as depicted in
Fig.~\ref{Fig1}(a). To break the rotational symmetry and achieve a
spin-dependent angular splitting, as shown in Fig.~\ref{Fig1}(b), we
deviate the propagation axis from the center of metasurface
[positions marked by green circles in
Figs.~\ref{Fig2}(b)-\ref{Fig2}(d)]. This scheme is similar to
scrubbing part of the inner structure of metasurface. Firstly, we
verify that the structure with rotational symmetry can covert a
circular polarized plane beam to its opposite circular polarization
state with a helical phase.

Figure~\ref{Fig3} shows the intensity distribution and interference
pattern for PB phase of the emerging beam when the metasurface
functioned with rotation symmetry. Metasurface corresponding to
Fig.~\ref{Fig2}(b) is selected as the example. The intensity located
in a single ring and a characteristic dark spot with zero intensity
in the center is a signature for the beam with helical phase
[Figs.~\ref{Fig3}(a)-\ref{Fig3}(b)]. The helical phase can be
confirmed by the interference with a reference beam shows fork
dislocation [Figs.~\ref{Fig3}(c)-\ref{Fig3}(d)]. By means of local
polarization transformation of the metasurface, it is possible to
convert a light beam with homogeneous elliptical polarization into a
vector beam with any desired polarization distribution in
higher-order Poincar\'{e} sphere~\cite{Liu2014}. Whereas, there is
no spin-dependent splitting in this evolution.

When the input and output states are orthogonal, we look forward to
detecting the cross-polarized components, hence, to register the
spin-dependent splitting~\cite{Bliokh2006}. The cross-polarized
field distributions are given in Fig.~\ref{Fig4}. Here, we use the
polarizer GLP1 to get the incident state as horizontal polarization
and the second polarizer GLP2 to obtain the vertical polarization.
The cross components suggest that photons with opposite helicities
accumulate at the opposite edges of the beam, and thereby provide a
direct evidence of spin-dependent splitting.

\begin{figure}
\includegraphics[width=8cm]{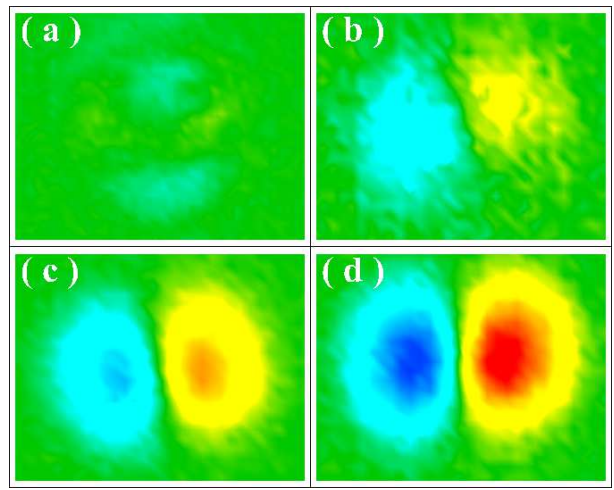}
\caption{\label{Fig5} (Color online) The $S_{3}$ parameter of the
photonic SHE: red and blue represent the right- and left-circular
polarizations, respectively. (a) The $S_{3}$ distribution for the
beam emerged from metasurface with rotational symmetry, which
corresponds to the structure marked by yellow circle in the
Fig.~\ref{Fig2}(b). [(b)-(d)] The $S_{3}$ results for metasurfaces
with rotational symmetry breaking, corresponding to green positions
in Figs.~\ref{Fig2}(b)-\ref{Fig2}(d). It is found that the splitting
is much more distinct and stronger with the increasing of rotation
rates. }
\end{figure}

To observe the photonic SHE intuitively and verify the
spin-dependent angular splitting, we also measure the Stokes
parameter $S_{3}$ in our experiment. It is known that $S_{3}$ is a
parameter to character the circular polarization
$S_{3}=\frac{I_{\sigma^{+}}-I_{\sigma^{-}}}{I_{\sigma^{+}}+I_{\sigma^{-}}}$~\cite{Born1997},
where $I_{\sigma^{+}}$ and $I_{\sigma^{-}}$ represent the intensity
of left- and right-circular polarization components, respectively.
By separately recording the intensities after orthogonal circular
polarizers, we calculate the Stokes parameter $S_{3}$ of the output
beam from the metasurface for each point in the beam cross-section.
Figure~\ref{Fig5} is the processed results of $S_{3}$, red and blue
represent the opposite spin states. Figure~\ref{Fig5}(a) is the
$S_{3}$ pattern for the incident position marked by the yellow
circle in Fig.~\ref{Fig2}(b), which is an example for the situations
with rotational symmetry [yellow circles in
Fig.~\ref{Fig2}(b)-\ref{Fig2}(d)]. It is found that there is no
splitting in these situations. Figures ~\ref{Fig5}(b)-\ref{Fig5}(d)
record the $S_{3}$ of emerging beams corresponding to the three
metasurfaces with rotational symmetry breaking and different
rotation rates. The incident positions are located at the same
distance off the center of the metasurfaces which are marked by the
green circles in Figs.~\ref{Fig2}(b)-\ref{Fig2}(d). It is found that
the splitting is much more distinctness and stronger with the
increasing of rotation rates, which proves that the spin-dependent
splitting is proportional to the rotation rate of the metasurface
agreeing well with our analysis.

As what we have discussed before, the angular splitting in real
position space would allow us to direct observe the photonic SHE.
However, observation of the spin-dependent splitting in experiment
depends both on the constructed geometric phase gradient and the
collimation of the beam. Our experiment results show that the
rotation rates of the three applied metasurface are not big enough
to show the intensity splitting. This is because that the splitting
angle constructed in our experiment is less than the far field
divergence angle~\cite{Born1997} of incident Gaussian beam
$\theta_t<\theta_d$, which can be written as
\begin{equation}
\theta_d=\frac{\lambda_{0}}{\pi w_{0}}\label{DA},
\end{equation}
where $w_{0}$ is the waist radius of Gaussian beam. The results in
Fig.~\ref{Fig5} demonstrate that the spin-dependent splitting can be
enhanced by increasing the rotation rate of metasurface structure.
Therefore, if we want to realize the spin-dependent splitting of
intensity, we should ensure that the rotation rate is large enough
to make the splitting angle much larger than the far field
divergence angle of the incident beam $\theta_t\gg\theta_d$.

\section{Conclusions}
In conclusion, we have demonstrated photonic SHE in metasurfaces
with rotational symmetry breaking. In Ref.~\cite{Yin2013}, this
effect was explained in the context of the spin-orbit interaction of
photons in a metasurface. Within their scheme, the metasurface
structure can be regarded as a simple case of rotational symmetry
breaking. From the viewpoint of geometric phases, the spin-dependent
splitting is perpendicular to the designed phase gradient, and
thereby is attributed to the RVB phase. In our scheme, the
spin-dependent splitting is parallel to the designed phase gradient,
and thereby is related to the PB phase. These interesting phenomena
in optical near field has also been discussed in surface plasmon
nanostructure with rotational symmetry breaking~\cite{Bliokh2008I}.
In our scheme, the metasurfaces were constructed by dielectric
nanograting and thereby the spin-dependent splitting can be detected
in far field with high transmission efficiency. We also note that
these results can be extrapolated to the electronic systems due to
the similarly geometrical phase roots~\cite{Karimi2012}. In
addition, our results can be directly applied to the vector beams
whose polarization takes a spatial rotation rate, namely exhibits an
inhomogeneous
polarization~\cite{Zhan2009,Holleczek2011,Milione2011,Milione2012,Ling2014I}.
The PB phase can be regarded as an intrinsic property of vector
beams, and spin-dependent splitting in momentum space would be
visualized when the rotation symmetry is broken~\cite{Ling2014II}.
We believe that these results may provide insights into the
fundamental properties of photonic SHE and Pancharatnam-Berry phase.

\begin{acknowledgements}
This research was partially supported by the National Natural
Science Foundation of China (Grants Nos. 61025024, 11274106, and
11347120), and the Scientific Research Fund of Hunan Provincial
Education Department of China (Grant No. 13B003).
\end{acknowledgements}

\end{document}